\documentclass[%
superscriptaddress,
preprint,
 amsmath,amssymb,
 aps,
]{revtex4-1}

\usepackage{graphicx}
\usepackage{dcolumn}
\usepackage{bm}
\usepackage{lineno}

\begin{document}


\title{Nanophotonic quantum storage at telecommunications wavelength}

\author{Ioana Craiciu}
	\affiliation{Kavli Nanoscience Institute and Thomas J. Watson, Sr., Laboratory of Applied Physics, California Institute of Technology, Pasadena, California 91125, USA}
	\affiliation{Institute for Quantum Information and Matter, California Institute of Technology, Pasadena, California 91125, USA}
\author{Mi Lei}
	\affiliation{Kavli Nanoscience Institute and Thomas J. Watson, Sr., Laboratory of Applied Physics, California Institute of Technology, Pasadena, California 91125, USA}
	\affiliation{Institute for Quantum Information and Matter, California Institute of Technology, Pasadena, California 91125, USA}
\author{Jake Rochman}
	\affiliation{Kavli Nanoscience Institute and Thomas J. Watson, Sr., Laboratory of Applied Physics, California Institute of Technology, Pasadena, California 91125, USA}
	\affiliation{Institute for Quantum Information and Matter, California Institute of Technology, Pasadena, California 91125, USA}
\author{Jonathan M. Kindem}
	\affiliation{Kavli Nanoscience Institute and Thomas J. Watson, Sr., Laboratory of Applied Physics, California Institute of Technology, Pasadena, California 91125, USA}
	\affiliation{Institute for Quantum Information and Matter, California Institute of Technology, Pasadena, California 91125, USA}
\author{$\mbox{John G. Bartholomew}$}
	\affiliation{Kavli Nanoscience Institute and Thomas J. Watson, Sr., Laboratory of Applied Physics, California Institute of Technology, Pasadena, California 91125, USA}
	\affiliation{Institute for Quantum Information and Matter, California Institute of Technology, Pasadena, California 91125, USA}
\author{Evan Miyazono}
	\affiliation{Kavli Nanoscience Institute and Thomas J. Watson, Sr., Laboratory of Applied Physics, California Institute of Technology, Pasadena, California 91125, USA}
	\affiliation{Institute for Quantum Information and Matter, California Institute of Technology, Pasadena, California 91125, USA}
\author{Tian Zhong}\altaffiliation[Currently at: ]{Institute of Molecular Engineering, University of Chicago, Chicago, Illinois 60637, USA}
	\affiliation{Kavli Nanoscience Institute and Thomas J. Watson, Sr., Laboratory of Applied Physics, California Institute of Technology, Pasadena, California 91125, USA}
	\affiliation{Institute for Quantum Information and Matter, California Institute of Technology, Pasadena, California 91125, USA}
\author{Neil Sinclair}
	 \affiliation{Division of Physics, Mathematics and Astronomy, California Institute of Technology, Pasadena, California 91125, USA}
	 \affiliation{Alliance for Quantum Technologies, California Institute of Technology, Pasadena, California 91125, USA}
\author{Andrei Faraon}
 	\email[Corresponding author: ]{faraon@caltech.edu}
	\affiliation{Kavli Nanoscience Institute and Thomas J. Watson, Sr., Laboratory of Applied Physics, California Institute of Technology, Pasadena, California 91125, USA}
	\affiliation{Institute for Quantum Information and Matter, California Institute of Technology, Pasadena, California 91125, USA}

\date{\today}

\begin{abstract}
Quantum memories for light are important components for future long distance quantum networks. We present on-chip quantum storage of telecommunications band light at the single photon level in an ensemble of erbium-167 ions in an yttrium orthosilicate photonic crystal nanobeam resonator. Storage times of up to $10 \: \mu$s are demonstrated using an all-optical atomic frequency comb protocol in a dilution refrigerator under a magnetic field of $\mbox{380 mT}$. We show this quantum storage platform to have high bandwidth, high fidelity, and multimode capacity, and we outline a path towards an efficient erbium-167 quantum memory for light. 
\end{abstract}

\pacs{Valid PACS appear here}
\maketitle

Optical quantum memories can aid processes involving the transfer of quantum information via photons, with applications in long distance quantum communication and quantum information processing \cite{Kimble2008,Heshami2016,Briegel1998,Kok2007,Monroe2014}. Rare-earth ions in crystals are a promising solid-state platform for optical quantum memories due to their long-lived optical and spin transitions that are highly coherent at cryogenic temperatures \cite{Thiel2011,Zhong2015}. Among rare-earth ions, only erbium has been shown to possess highly coherent optical transitions in the telecommunications C-band, which allows for integration of memory systems with low loss optical fibers and integrated silicon photonics \cite{Miyazono2017}.

Fixed delay quantum storage for less than $\mbox{50 ns}$ at telecommunications wavelengths has been demonstrated in erbium-doped fibers \cite{Saglamyurek2015}, and lithium niobate waveguides \cite{Askarani2018} at efficiencies approaching $1\%$. The protocol used in both cases, the atomic frequency comb (AFC), requires spectrally selective optical pumping \cite{Afzelius2009}. The storage efficiencies in these works were limited in part by the lack of suitable long-lived shelving states in the erbium ions in these hosts. Moving to isotopically purified erbium-167 in a yttrium orthosilicate host (YSO) offers the prospect of long-lived shelving states in the form of hyperfine levels \cite{Rancic2018}. While optical storage has been realized in erbium-doped YSO \cite{Lauritzen2010,Lauritzen2011,Dajczgewand2014}, including efficiencies approaching $50\%$ at storage times of $16 \: \mu$s (revival of silenced echo protocol \cite{Dajczgewand2014}) quantum storage has yet to be demonstrated in this material.

In this work, we demonstrate on-chip quantum storage of telecommunications light at the single photon level. We used a nanophotonic crystal cavity milled directly in $^{167}$Er$^{3+}$ doped YSO ($^{167}$Er$^{3+}$:YSO) to couple to an ensemble of erbium ions and realize quantum storage using the AFC protocol \cite{Afzelius2009}. The cavity increased the absorption of light by the ion ensemble, allowing on-chip implementation of the memory protocol \cite{Afzelius2010}. By working in a dilution refrigerator and using permanent magnets to apply a field of $\mbox{380 mT}$, we accessed a regime in which the ions have optical coherence times of $\sim 150 \: \mu$s and long lived spin states to allow spectral tailoring. For the shortest measured storage time of $\mbox{165 ns}$, we achieved an efficiency of $0.2 \%$, with lower efficiencies for longer storage times, up to $10 \: \mu$s. We demonstrated storage of multiple temporal modes and measured a high fidelity of storage, exceeding the classical limit. Lastly, we identified the limits on the storage efficiency and proposed avenues for overcoming them to achieve an efficient $^{167}$Er$^{3+}$ quantum memory for light.

Memories using spectral tailoring such as the AFC protocol require a long-lived level within the optical ground state manifold, where population can be shelved. Hyperfine levels in the optical ground state in $^{167}$Er$^{3+}$:YSO have been shown to have long lifetimes at 1.4 K and a magnetic field of $\mbox{7 T} $\cite{Rancic2018}. In general, these levels can be long-lived when the erbium electron spin is frozen, which occurs when $\hbar \omega_e \gg k_B T$, where $\omega_e$ is the electron Zeeman splitting \cite{Rancic2018}. In this work, we satisfied this inequality by using a moderate magnetic field of $\mbox{380 mT}$ parallel to the D$_1$ axis of the crystal $(\omega_e = 2 \pi \times 80 \: \textrm{GHz})$ and a nanobeam temperature of $\mbox{$\sim$400 mK}$. The nanobeam temperature was estimated via the electron spin population at a lower magnetic field \cite{SOM,Kukharchyk2018}. The hyperfine lifetime in the bulk crystal was measured to be $\mbox{29 minutes}$, enabling the long-lived, spectrally selective optical pumping required for the AFC protocol (see supplementary material \cite{SOM}).

\begin{figure*}
  \centering
      \includegraphics[width=6.6in]{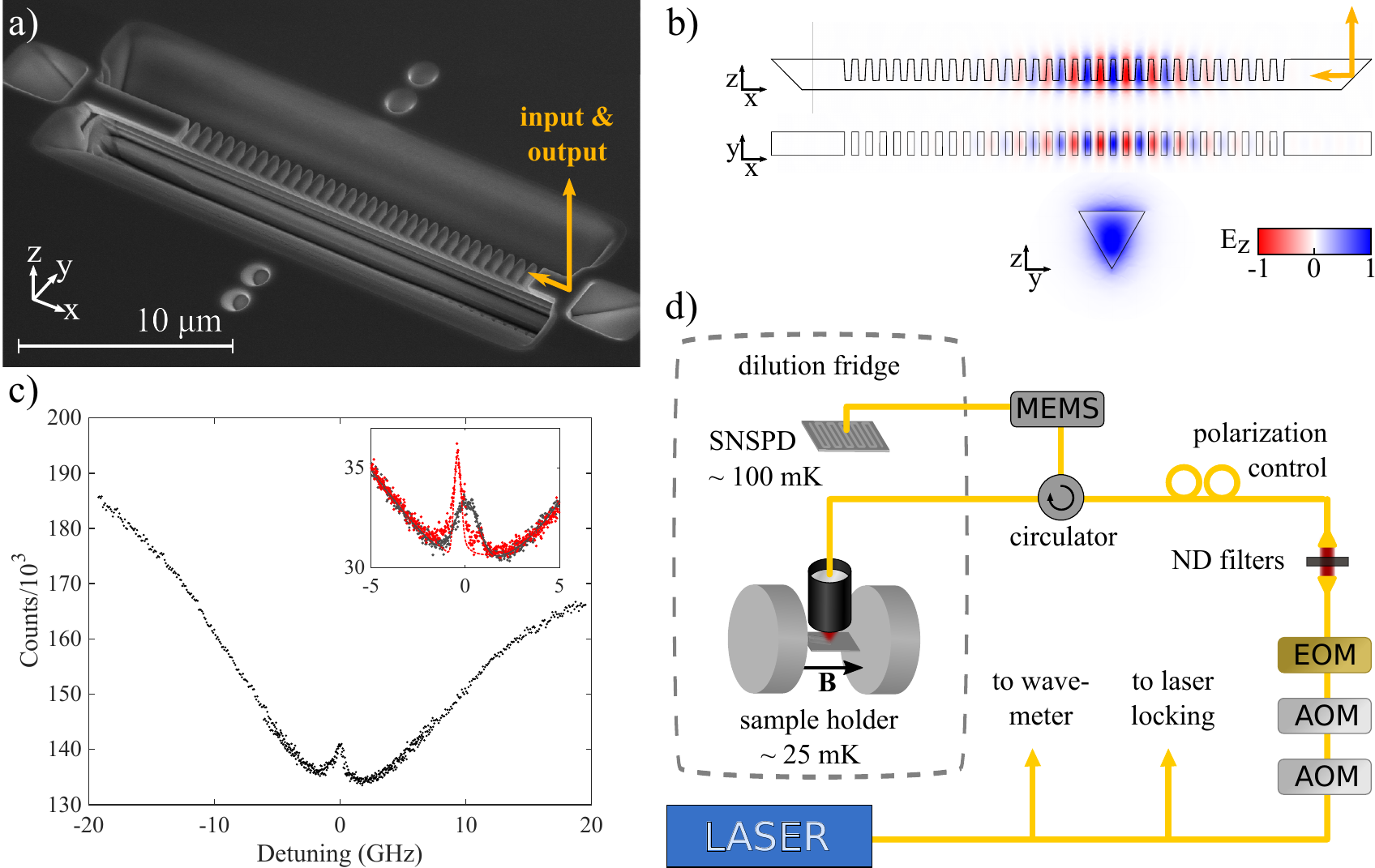}
  \caption{\label{fig1} (a) Scanning electron micrograph of the triangular nanobeam resonator, showing input/output coupling through a $45^\circ$ angled slot coupler. (b) Simulation of the TM cavity mode. Red-blue color gradient indicates the electric field component normal to the surface, $E_z$; black outline indicates YSO-air interface; yellow arrow indicates coupling. (c) Reflection spectrum of cavity when tuned on resonance to the $\mbox{1539 nm}$ $^{167}$Er$^{3+}$:YSO transition. Detuning is measured from $\mbox{194816 GHz $\pm$ 2 GHz}$. Inset shows a close-up of ion coupling before (black) and after (red) partial hyperfine initialization. Circles are data points; solid black and dashed red lines are fits to theory (see main text for details). (d) Schematic of setup. Light from an external-cavity diode laser was directed through two acousto-optical modulators (AOMs) for pulse shaping, and an electro-optic phase modulator (EOM) to add strong sidebands for hyperfine initialization. Neutral density (ND) filters and polarization paddles provided attenuation and polarization control, respectively. A circulator directed light to the $^{167}$Er$^{3+}$:YSO crystal located inside a dilution refrigerator, thermally linked to the $\mbox{25 mK}$ stage. An aspheric lens pair focused light from an optical fiber onto the angled coupler of the resonator. Light from the resonator was directed by the circulator onto a superconducting nanowire single photon detector (SNSPD) at $\sim 100$ mK. A micro electro-mechanical switch (MEMS) prevented strong initialization pulses from reaching the SNSPD. A magnetic field $\mbox{$\mathbf{B} = 380$ mT}$ was applied to the sample using two cylindrical permanent magnets. More details in the supplementary material \cite{SOM}.}
\end{figure*}

Figures \ref{fig1}a-c describe the nanoresonator used in this experiment. A triangular nanobeam photonic crystal cavity \cite{Zhong2016} was milled in a YSO crystal doped with isotopically purified $^{167}$Er$^{3+}$ ($92 \%$ purity) at a nominal concentration of $\mbox{50 ppm}$. The width of the nanobeam was $1.5 \: \mu$m, and the length $\sim 20 \: \mu$m. The slots in the nanobeam created a photonic crystal bandgap, and the periodic pattern ($\mbox{lattice constant = 590 nm}$, $\mbox{groove width = 450 nm}$) was modified quadratically in the center to create a cavity mode. Figure \ref{fig1}a shows a scanning electron micrograph of the nanobeam and Fig$.$ \ref{fig1}b shows a finite element analysis simulation of the TM cavity mode.

The coherence time of the $\mbox{1539 nm}$ optical transition, which provides an upper bound on all-optical storage time, was measured to be $149 \: \mu\mathrm{s} \pm 4 \: \mu\mathrm{s}$ in the nanobeam. The bulk optical coherence time of this transition under similar cooling conditions was measured to be $760 \: \mu\mathrm{s} \pm 41 \: \mu\mathrm{s}$. The reduction in coherence time as measured in the nanobeam is likely caused by a combination of higher temperature in the nanobeam during measurement and the impact of the focused ion beam milling process \cite{SOM}. We note that the fabrication method has not significantly impacted the coherence properties of ions in similar devices \cite{Zhong2015a,Zhong2017}, although the bulk crystal coherence times measured in those works were also lower, preventing a direct comparison. The optical coherence time did not limit the storage time achieved in this work.

Figure \ref{fig1}d shows a schematic of the optical testing setup. Figure \ref{fig1}c shows the reflection spectrum of the nanobeam cavity, which has a measured loaded quality factor of $7 \times 10^3$. The cavity was tuned onto resonance with the $\mbox{1539 nm}$ transition of the $^{167}$Er$^{3+}$ ions by freezing nitrogen gas onto the nanobeam at cryogenic temperatures \cite{Mosor2005}. The coupling of the ensemble of ions to the cavity is seen as a peak in the cavity reflection dip. The inset shows a close-up of the ion-cavity coupling (in black) and a fit to theory \cite{Diniz2011}. The ensemble cooperativity was estimated from the fit to this curve to be 0.1 \cite{SOM}. For high efficiency storage using ions coupled to a cavity, the ensemble cooperativity (after spectral tailoring) should equal one \cite{Afzelius2010,Moiseev2010}. An increased ensemble cooperativity of 0.3, as shown in red in the inset of Fig$.$ \ref{fig1}c, was obtained using a partial hyperfine initialization procedure. The spectral density of erbium ions in the center of the inhomogeneous line (zero detuning) was increased from its thermal equilibrium value by sweeping the laser frequency between 350 MHz and 820 MHz on both sides of the inhomogeneous line. This procedure can be optimized to further increase the ensemble cooperativity, but was limited here by the relatively weak magnetic field, which allows only partial hyperfine initialization \cite{SOM}. At $\mbox{7 T}$, initialization into one hyperfine state with an efficiency of $95\%$ has been demonstrated \cite{Rancic2018}.

\begin{figure}
  \centering
      \includegraphics[width=3.5in]{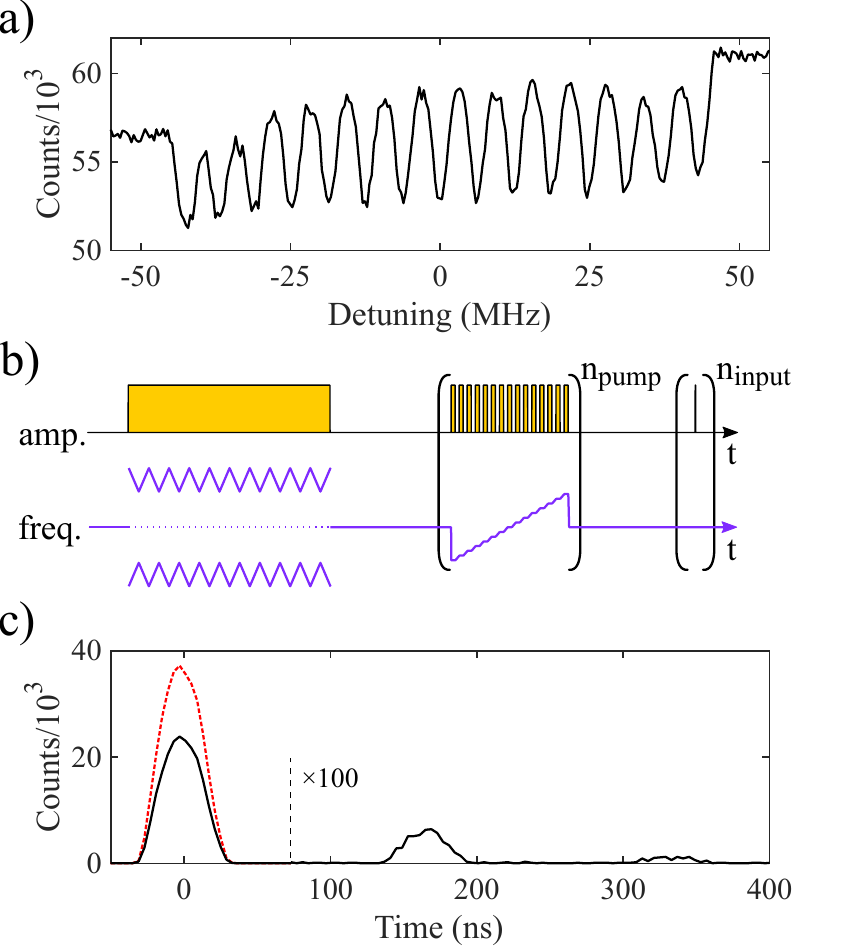}
  \caption{\label{fig2} (Color online) AFC experiment in the nanobeam cavity. (a) A section of the resonator reflection spectrum, showing an atomic frequency comb in the center of the inhomogeneously broadened $^{167}$Er$^{3+}$ transition. Detuning is measured from $\mbox{194814.2 GHz $\pm$ 0.1 GHz}$. The apparent slope of the comb is due to its center frequency not being precisely aligned to the cavity resonance, leading to a dispersive shape. (b) Schematic of AFC pulse sequence showing amplitude (yellow) and frequency (purple) modulation of the laser. The pulse widths and heights are not drawn to scale. The details of the sequence are described in the main text. (c) AFC storage: the input pulse (red dashed line) was partially absorbed by the comb and an output (stored) pulse was emitted at time $\mbox{$1/\Delta =165$ ns}$. The black line shows the partially reflected input pulse and the output pulse intensity $(\times 100)$. A smaller second output pulse is seen at 330 ns.}
\end{figure}

The nanobeam device was used to demonstrate quantum optical storage using the AFC protocol \cite{Afzelius2009}. In this protocol, a pulse of light that is absorbed by an AFC with an inter-tooth spacing of $\Delta$ is stored for $t=1/\Delta$. Frequency selective optical pumping was used to create a comb within the inhomogeneous linewidth, as shown in Fig$.$ \ref{fig2}a. Figure \ref{fig2}b shows a schematic of the protocol. First, a long pulse with strong frequency modulated sidebands was used for partial hyperfine initialization. The next 15 pulses, repeated $n_\mathrm{pump} = 20$ times, created the comb: the laser frequency was swept through 15 values, separated by $\mbox{$\Delta = 6.1$ MHz}$, to optically pump away ions and create 15 spectral transparencies. The following $n_\mathrm{input} = 60$ pulses were zero-detuning weak coherent states which were stored in the frequency comb. The full experiment was repeated $\sim 10^4$ times. As shown in Fig$.$ \ref{fig2}c, $\mbox{60 ns}$ wide pulses with an average photon number of $\bar n = 0.60 \pm 0.09$ were stored for $\mbox{165 ns}$ with an efficiency of $0.2 \%$. Despite the partial initialization, the storage efficiency was limited by the ensemble cooperativity of the device \cite{SOM}. 

\begin{figure}
  \centering
      \includegraphics[width=3.5in]{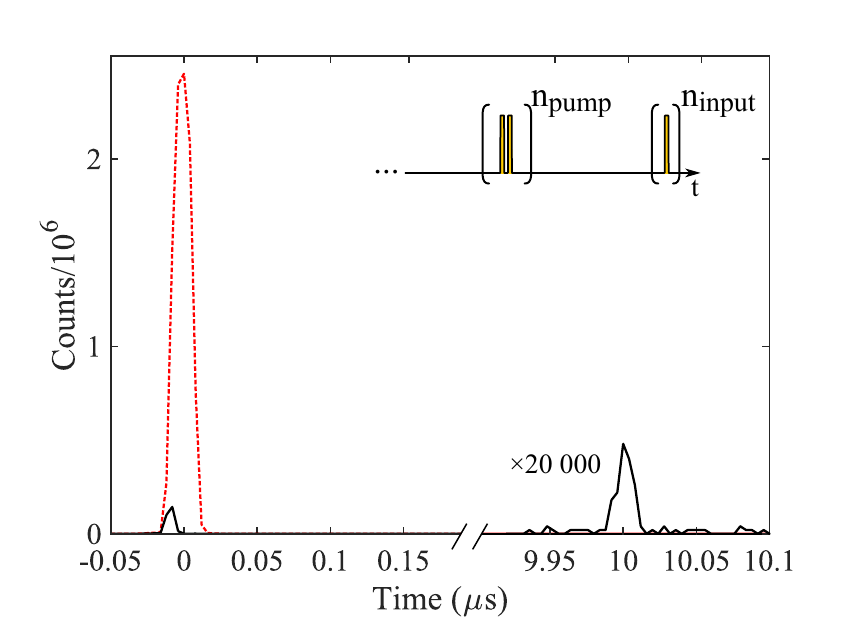}
  \caption{\label{fig3} (Color online) AFC storage for $\mbox{$10 \: \mu$s}$ in the nanobeam resonator.  Red dashed line shows the input pulse. Black line shows the partially reflected input pulse and the output pulse $(\times 20\,000)$. The reflected input pulse appears small due to detector saturation. Inset shows a schematic of the pulse sequence following hyperfine initialization. Pairs of comb preparation pulses $\mbox{$10\: \mu$s}$ apart were repeated $n_\mathrm{pump} = 10\,000$ times, followed by input pulses $\mbox{20 ns}$ wide, repeated $n_\mathrm{input} = 10$ times.}
\end{figure}

Coherent pulses could be stored in the device for up to $10 \: \mu$s, although with a lower efficiency of $10^{-5}$, as shown in Fig$.$ \ref{fig3}. For this experiment, as for all storage times longer than $\mbox{165 ns}$, we used an accumulated AFC method \cite{SOM,DeRiedmatten2008} to create the comb. As shown in the inset of Fig$.$ \ref{fig3}, weak pairs of pulses separated by $t_\mathrm{storage} = 10 \: \mu$s were repeatedly sent into the cavity. The Fourier transform of each pulse pair is a frequency comb, which imprinted onto the $^{167}$Er$^{3+}$ inhomogeneous line to create the AFC. The efficiency at this long storage time was limited by laser frequency jitter and by superhyperfine coupling to the yttrium ions in YSO. Superhyperfine coupling limits the narrowest spectral feature to $\sim 1$ MHz \cite{Guillot-Noel2007,Car2018}. Since this exceeds the period of the comb needed for this storage time ($\Delta = 1/t_\mathrm{storage} = 0.1$ MHz), the resulting AFC will have a lower contrast, leading to lower storage efficiency \cite{SOM}.

\begin{figure}[b!]
  \centering
      \includegraphics[width=3.5in]{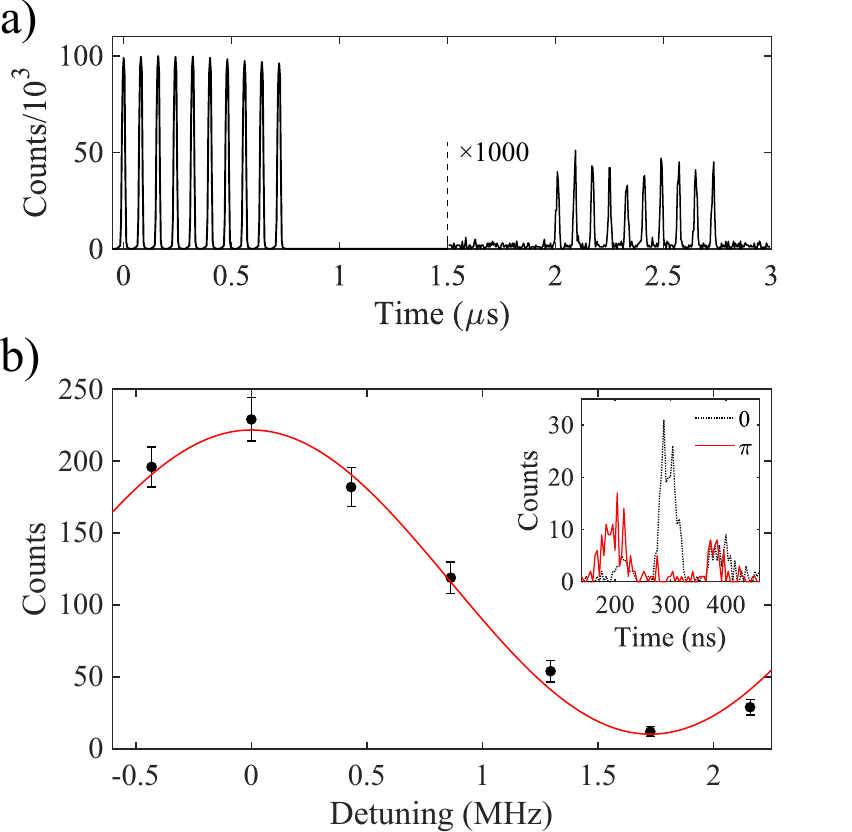}
  \caption{\label{fig4} (Color online) Multimode and coherent storage in the nanobeam resonator. (a) Storage of multiple temporal modes: ten $\mbox{20 ns}$ wide input pulses (reflection off cavity shown) and the corresponding 10 output pulses from a 500 kHz AFC $(\times 1000)$. (b) Visbility curve acquired using double comb experiment, with $\Delta_1 = 5.2$ MHz, $\Delta_2 = 3.4$ MHz, $\delta_1 = 0$ MHz. The  detuning of the second comb was swept from $\delta_2 = -0.2$ MHz to $\delta_2 = 2.2$ MHz, and the intensity of the two central overlapping output pulses was measured. Black circles show the sum of counts in the overlapping pulse region with $\sqrt{N_\mathrm{counts}}$ uncertainty bars. Red line shows a least squares fit to a sinusoid. Inset shows the four output pulses (middle two overlapping) in the case of the maximally constructive (dashed black line) and maximally destructive (solid red line) interference.}
\end{figure}

The AFC protocol is capable of storing multiple temporal modes \cite{Afzelius2009}. Ten coherent pulses were stored in this device, as shown in Fig$.$ \ref{fig4}a. Multiplexing in frequency is also possible \cite{Sinclair2014}. The AFC comb in Figure \ref{fig2}a has a bandwidth of $\sim 90$ MHz (see Fig$.$ \ref{fig2}a) which can accommodate storage in multiple frequency modes. An inhomogeneous linewidth of 150 MHz limits the bandwidth of storage in this system. Although there exist methods to increase this linewidth \cite{Welinski2017}, the bandwidth cannot be increased much further before being limited by overlapping optical transitions from other hyperfine levels.

In quantum storage protocols, the phase of the stored state must be preserved. A double AFC was used as an interferometer in order to characterize the coherence of the storage process \cite{DeRiedmatten2008}. Two overlapping AFCs with tooth spacing $\Delta_1, \Delta_2$ and with frequency detuning $\delta_1, \delta_2$, were created, so that each input pulse was mapped to two output pulses at times $1/\Delta_1, 1/\Delta_2$ and with a relative phase $\phi_\mathrm{rel}= 2 \pi \left( \frac{\delta_2}{\Delta_2}-\frac{\delta_1}{\Delta_1}\right)$\cite{Afzelius2009}.  An input state encoded into two pulses, $| \psi_\mathrm{in} \rangle = \frac{1}{\sqrt{2}} \left( | \mathrm{early} \rangle + | \mathrm{late} \rangle \right)$, was therefore mapped to a total of four output pulses. By appropriately selecting the time interval between the early and late input pulses, two of the four output pulses were made to overlap and either constructively or destructively interfere, depending on $\phi_\mathrm{rel}$ (see inset of Fig. \ref{fig4}b). Using an input state with mean photon number $\bar n = 0.6 \pm 0.09$, and sweeping $\phi_\mathrm{rel}$ via the detuning $\delta_2$, the interference fringe shown in Fig$.$ \ref{fig4}b was obtained (see caption for details). The measured visibility of $91.2\% \pm 3.4\%$ demonstrates the high degree of coherence of this on-chip storage process. The visibility was limited by the 12 counts in the total destructive interference case $\left( \delta_2 = \frac{\Delta_2}{2} \rightarrow \phi_\mathrm{rel} = \pi \right)$. This was due in part to imperfect cancellation of the two overlapping output pulses, resulting from the slightly different efficiencies of storage in the two AFCs, and in part to a dark count rate of 18.5 Hz, leading to a baseline of 7 counts. The dark-count-subtracted visibility is $97.0\% \pm 3.6\%$. 

The double comb method was also used to estimate a lower bound for the fidelity of storing single photon time bin states, $F^{(n=1)}$. The fidelity of storage was measured for four input states $|\mathrm{early}\rangle$, $|\mathrm{late}\rangle$,$\frac{|\mathrm{early}\rangle+|\mathrm{late}\rangle}{\sqrt{2}}$ and $\frac{|\mathrm{early}\rangle-|\mathrm{late}\rangle}{\sqrt{2}}$, using two mean photon numbers $\bar n = 0.30$ and $\bar n = 0.60$. With these values, the decoy state method \cite{SOM,Sinclair2014,Ma2005} was used to calculate a bound on the fidelity for storing single photon states $F^{(n=1)}\geq 93.7 \% \pm 2.4\%$, which exceeds the classical limit of $F=2/3$. Similar to the case for visibility discussed above, the measured fidelity was limited in part by dark counts and in part by the double comb protocol being an imperfect interferometer \cite{SOM}. Although the pulses stored in this experiment were weak coherent pulses, single \cite{Zargaleh2018,Wang2018,Rielander2016,Dibos2018,Paul2017,Atzeni2018} and entangled photon \cite{,Niizeki2018,Zhou2018} sources are available at telecommunications wavelengths. Notably, sources in the telecommunications C-band \cite{Dibos2018,Paul2017,Atzeni2018,Niizeki2018} including individual erbium-ion emitters \cite{Dibos2018}, are compatible with this device. Storing photons of any frequency is enabled by quantum frequency conversion \cite{Dreau2018}. The bandwidth of compatible photons is bounded by the AFC periodicity of $\mbox{$\gtrsim$ 1 MHz}$ and the $\mbox{$\lesssim$ 100 MHz}$ bandwidth of the AFC.

While the storage presented here was limited in efficiency, a nanophotonic cavity coupled to $^{167}$Er$^{3+}$ ions in YSO promises to be an efficient quantum storage system. The main limitations to the storage efficiency in this work were a low ensemble cooperativity of 0.3 and loss from the optical nanobeam cavity. The cooperativity can be increased using higher $^{167}$Er$^{3+}$ doping and better hyperfine initialization, which would require increasing the applied magnetic field or changing its angle \cite{SOM,Rancic2018}. A higher intrinsic quality factor resonator would serve to both increase cooperativity and decrease cavity loss. For example, using a YSO crystal with 200 ppm $^{167}$Er$^{3+}$ doping, optimal hyperfine initialization, and a resonator with an intrinsic quality factor of 2 million, the theoretical efficiency of the AFC quantum storage is $90\%$ (see supplementary material for analysis \cite{SOM}). Mature silicon nanofabrication technology can be leveraged to achieve this goal by using a silicon resonator evanescently coupled to $^{167}$Er$^{3+}$ ions in YSO \cite{Dibos2018,Miyazono2017}. With this efficiency level and a storage time of $10 \: \mu$s, the device would outperform a delay line composed of standard telecommunications fiber \cite{Cho2016}, an important benchmark on the way to achieving a quantum memory suitable for scalable quantum networks.

With the optical AFC protocol alone, it will be difficult to achieve efficient storage for this duration due to superhyperfine coupling. However, the AFC spin-wave protocol, where the stored information is reversibly transferred from the optical to the hyperfine manifold \cite{Afzelius2009}, would enable even longer storage without the same requirements for narrow spectral features, as well as enabling on-demand recall. The availability of hyperfine states with coherence times exceeding $\mbox{1 second}$ \cite{Rancic2018} make $^{167}$Er$^{3+}$:YSO a promising system for spin-wave storage.  

In conclusion, we have demonstrated on-chip quantum storage of telecommunications band light at the single photon level. The storage had a bandwidth of $\mbox{$\sim 90$ MHz}$, and a storage fidelity for single photon states of at least $93.7 \% \pm 2.4\%$. $^{167}$Er$^{3+}$:YSO at temperatures of $\sim$ 400 mK and moderate magnetic field was shown to be a promising material for AFC quantum memories. A clear path exists for creating a high efficiency quantum memory using this material and a nanoscale resonator. 

\begin{acknowledgments}
This work was supported by AFOSR Young Investigator Award [FA9550-15-1-0252], AFOSR Grant [FA9550-18-1-0374], and the National Science Foundation [EFRI 1741707]. I.C and J.R. acknowledge the support from the Natural Sciences and Engineering Research Council of Canada (NSERC) [PGSD2-502755-2017, PGSD3-502844-2017]. J.G.B. acknowledges the support of the American Australian Association's Northrop Grumman Fellowship. N.S. acknowledges funding by the Alliance for Quantum Technologies' Intelligent Quantum Networks and Technologies (INQNET) research program.
\end{acknowledgments}

\bibliography{Craiciu_2019_Nanophotonic_quantum_storage_at_telecommunications_wavelength_bibliography_arxiv}
\bibliographystyle{apsrev4-1}
\end{document}